\documentclass[sort&compress,
     final           
  4apaper]
  {aipproc}

\layoutstyle{8x11single}

\usepackage{bm}
\usepackage{amsmath}
\usepackage{amssymb}

\newcommand\beq{\begin{equation}}
\newcommand\eeq{\end{equation}}
\newcommand\beqa{\begin{eqnarray}}
\newcommand\eeqa{\end{eqnarray}}

\newcommand{\rr}{\mathbf{r}}
\newcommand{\cc}{\mathbf{v}}
\newcommand{\VV}{\mathbf{V}}

\begin{document}

\title{Velocity Distribution and Cumulants in the Unsteady Uniform Longitudinal Flow of a Granular Gas}


\author{Antonio Astillero}{
  address={Departamento de Tecnolog\'ia de Computadores y Comunicaciones, Universidad de Extremadura, E-06800 M\'erida, Spain} }

\author{Andr\'es Santos}{
  address={Departamento de F\'{\i}sica, Universidad de
Extremadura, E-06071 Badajoz, Spain} }

\begin{abstract}
The  uniform longitudinal flow is characterized by a linear longitudinal velocity field
$u_x(x,t)=a(t)x$, where $a(t)={a_0}/({1+a_0t})$ is the strain rate, a uniform  density
$n(t)\propto a(t)$,  and a uniform granular temperature $T(t)$. Direct simulation Monte Carlo solutions of the Boltzmann equation for inelastic hard spheres are presented for three (one positive and two negative) representative values  of the initial strain rate $a_0$.
Starting from different initial conditions, the temporal evolution of the reduced strain rate $a^*\propto a_0/\sqrt{T}$, the non-Newtonian viscosity, the second and third velocity cumulants, and three independent marginal distribution functions has been recorded. Elimination of time in favor of the reduced strain rate $a^*$ shows that, after a few collisions per particle,  different initial states are attracted to  common ``hydrodynamic'' curves. Strong deviations from Maxwellian properties are observed from the analysis of the cumulants and the marginal distributions.
\end{abstract}

\maketitle


\section{Introduction}

The  dynamical properties of granular gases are  in general much more complex than those of  conventional molecular gases due to several causes such as, for instance, collisional inelasticity, frictional effects, polydispersity, non-sphericity, or influence of the interstitial fluid. In order to isolate the first effect from the other ones, a favorite  model of a granular gas consists of an ensemble of identical, smooth, inelastic  hard spheres with a constant coefficient of normal restitution $\alpha$ \cite{C90,D00,G03,BP04}. In the dilute regime,  a kinetic theory approach based on the Boltzmann  equation
\beq
\partial_t f(\rr,\cc, t)+\cc\cdot \nabla f(\rr,\cc,t)=J[\cc|f,f]
\label{2.1}
\eeq
has proven to be very powerful. In Eq.\ \eqref{2.1}, $f(\rr,\cc, t)$ is the one-body velocity distribution function and  $J[\cc|f,f]$ is the Boltzmann operator for inelastic collisions \cite{BDS97}.

In this work, we consider this simple model of a  granular gas under conditions of
uniform longitudinal flow (ULF) and analyze the temporal evolution of the velocity distribution function and its first few moments in the \emph{hydrodynamic} stage \cite{AS12}, i.e., once the kinetic stage (strongly sensitive to the initial state) has decayed.
The ULF \cite{AS12,GK96,KDN97,UP98,KDN98,UG99,S00a,S08a,S09} is characterized
by a linear longitudinal velocity field, a uniform density, and a uniform granular temperature $T(t)$:
\begin{equation}
u_{x}(x,t)=a(t)x, \quad n(t)=\frac{n_0}{a_0} a(t), \quad a(t)=\frac{a_{0}}{1+a_{0}t},
\label{velocity_field}
\end{equation}
where $a(t)$ is the strain rate.  It is important to note that the constant $a_0$ (initial strain rate) can be either positive (expansion of the gas) or negative (compression of the gas). The ULF is schematically depicted in Fig.\ \ref{ulf_sketch}.

The energy balance equation is given by \cite{AS12,S09}
\begin{equation}
\partial_{t}T(t)=-\frac{2}{3}a(t)T_{x}(t)-
\zeta(t)T(t),
\label{energy_balance}
\end{equation}
where $T_{x}=P_{xx}/n$ is the \emph{anisotropic} temperature along the $x$ direction (related to the normal stress $P_{xx}$) and $\zeta(t)$ is the cooling rate, which vanishes for elastic collisions ($\alpha=1$). If $a(t)>0$ (expansion), both terms on the right-hand side of Eq.\ \eqref{energy_balance} are negative and thus the granular gas monotonically loses kinetic energy, i.e., $\partial_t T(t)<0$. On the other hand, if $a(t)<0$ (compression) the  viscous heating term $\frac{2}{3}|a(t)|T_{x}(t)$ competes with the inelastic cooling term $\zeta(t)T(t)$ and, depending on the initial state, the temperature either grows or decays until a steady state is eventually reached.

 The relevant control parameter of
the problem is the \emph{reduced} strain rate (which plays the role of the Knudsen number)
\begin{equation}
a^{*}(t)=\frac{a(t)}{\nu(t)}\propto \frac{a_0}{\sqrt{T(t)}},
\label{longitudinal_rate}
\eeq
where $\nu(t)\propto n(t)\sqrt{T(t)}$ is an effective collision
frequency. A convenient choice is
\beq
\nu(t)=\frac{n(t)T(t)}{\eta_{\text{NS}}(t)}= \frac{1}{1.016}\frac{16\sqrt{\pi}}{5}\sigma^2n(t)\sqrt{\frac{T(t)}{m}}.
\label{nuB}
\eeq
Here, $\eta_{\text{NS}}$ is the  Navier--Stokes (NS) shear viscosity of a gas of \emph{elastic} hard spheres \cite{CC70},
$\sigma$ and $m$ being the diameter and mas  of a sphere, respectively. Obviously, $|a^*(t)|$ increases (decreases) with time in cooling (heating) situations and reaches a stationary value $a_s^*<0$ only if $a_0<0$.

In Ref.\ \cite{AS12} we presented direct simulation Monte Carlo (DSMC) results for the  evolution of the (reduced) strain rate $a^*(t)$ and the (reduced) non-Newtonian viscosity
\beq
\eta^*(t)=\frac{3}{4}\frac{1}{a^*(t)}\left[1-\frac{T_{x}(t)}{T(t)}\right]
\label{eta}
\eeq
for a wide ensemble of initial conditions. Parametric plots of $\eta^*(t)$ versus $a^*(t)$ showed that, after a first (kinetic) stage lasting a few collisions per particle, the system reached a second (hydrodynamic) stage where, regardless of the details of the initial state, the curves were attracted to a common smooth ``universal'' curve $\eta^*(a^*)$.
On the other hand, the viscosity $\eta^*$ involves second-order moments only and thus the possibility that the underlying full velocity distribution function might still be affected by the initial preparation, even when $\eta^*(a^*)$ exhibits a hydrodynamic behavior, was not addressed in Ref.\ \cite{AS12}.
The aim of the present work is to clarify this issue by extending that analysis to higher-order moments, namely the second ($a_2$) and third ($a_3$) cumulants, and to the velocity distribution itself.

\begin{figure}
\scalebox{1.2}[1.2]{\includegraphics{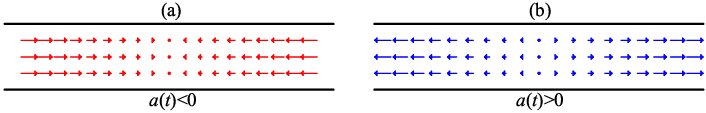}}
\caption{Sketch of the ULF for (a) $a(t)<0$ and (b) $a(t)>0$.}
\label{ulf_sketch}
\end{figure}

\section{Uniform longitudinal flow}
As said above, the ULF is defined by the macroscopic fields \eqref{velocity_field}, together with $\nabla T=0$ and the balance equation \eqref{energy_balance}. At a more basic level, the velocity distribution function $f(\rr,\cc,t)$ becomes spatially uniform when the velocities are referred to
a Lagrangian frame moving with the flow, i.e.,
\beq
f(\rr,\cc,t)=n(t)\rho({\bf V},t), \quad {\bf V}\equiv {\bf v}-{\bf u}(x,t),
\label{rho}
\eeq
where  $\rho({\bf V},t)$ is the probability density function and $\mathbf{V}$ is the peculiar velocity. After simple algebra, Eq.\ \eqref{2.1} can be
rewritten as \cite{AS12,S00a}
\beq
\partial_\tau \rho(\mathbf{V}, \tau)-a_0\frac{\partial}{\partial V_x}\left[V_x \rho(\mathbf{V}, \tau)\right]=n_0J[\mathbf{V}|\rho,\rho],\quad \tau\equiv \frac{\ln(1+a_0 t)}{a_0}.
\label{2.3}
\eeq
Equation (\ref{2.3}) shows that the original ULF problem can be mapped onto the equivalent problem of a {\em
uniform\/} gas with a velocity distribution $n_0\rho(\mathbf{V}, \tau)$ and
subject to the action of a non-conservative force $-m a_0V_x
\widehat{\bf x}$. Moreover, in the mapped problem the temporal evolution is monitored by the scaled variable $\tau=\int_0^t d  t'
n(t')/n_0$, which is unbounded even if $a_0<0$ since in that case $\tau\to
\infty$ when $t\to |a_0|^{-1}$. In terms of $\rho(\mathbf{V}, \tau)$, the longitudinal temperature $T_x$ and the average temperature $T$ are defined as
\beq
T_x(\tau)=m\langle V_x^2\rangle,\quad T(\tau)=\frac{m}{3}\langle V^2\rangle, \quad \langle \psi(\mathbf{V})\rangle\equiv \int d\mathbf{V}\,
\psi(\mathbf{V})\rho(\mathbf{V}, \tau).
\label{Tx}
\eeq
The transverse temperatures $T_y$ and $T_z$ can be defined similarly to $T_x$. Note that $T=\frac{1}{3}(T_x+T_y+T_z)$. The energy balance equation \eqref{energy_balance} can be equivalently written in the form
\begin{equation}
\partial_{\tau}T(\tau)=-\frac{2}{3}a_0T_{x}(\tau)-
\zeta_0(\tau)T(\tau), \quad \zeta_0(\tau)=-\frac{mn_0}{3T(\tau)}\int d\mathbf{V}\,
V^2J[\mathbf{V}|\rho,\rho],
\label{energy_balance2}
\end{equation}
where the cooling rate $\zeta_0(\tau)$ in the mapped problem is related to the cooling rate $\zeta(t)$ of the original problem by $\zeta_0(\tau)=n_0\zeta(t)/n(t)$.
In the mapped description, the roles of strain rate and collision frequency are played by $a_0$ and $\nu_0(\tau)=n_0\nu(t)/n(t)$, respectively. Therefore, the reduced strain rate \eqref{longitudinal_rate} remains the same in both descriptions.

Apart from the second-order moments $T_x(\tau)$ and $T(\tau)$, higher-order moments provide information about the distribution $\rho(\mathbf{V},\tau)$. In particular, deviations from a Maxwellian can be characterized by the second and third cumulants defined as
\begin{equation}
a_2(\tau)=\frac{\langle V^4\rangle}{15\left[T(\tau)/m\right]^2}-1,\quad a_3(\tau)=-\frac{\langle V^6\rangle}{105\left[T(\tau)/m\right]^3}+1+3a_2(\tau).
\label{cumulants}
\end{equation}

In general, the probability distribution function $\rho(\mathbf{V},\tau)$ depends on the three components of $\mathbf{V}$. It is then convenient to introduce the \emph{marginal} distributions
\begin{equation}
\rho_x(V_x,\tau)=\int_{-\infty}^\infty d
V_y\int_{-\infty}^\infty d V_z\,\rho(\mathbf{V},\tau),\quad
\rho_y(V_y,\tau)=\int_{-\infty}^\infty d
V_x\int_{-\infty}^\infty d V_z\,\rho(\mathbf{V},\tau),
\label{gx_mas_menos}
\end{equation}
\begin{equation}
P(V,\tau)=V^2\int d \widehat{\mathbf{V}}\,\rho(\mathbf{V},\tau).
\label{f_modulo}
\end{equation}
While the functions $\rho_x(V_x,\tau)$ and $\rho_y(V_y,\tau)$ provide information about the anisotropy of the state,
$P(V,\tau)$ is the probability distribution function of the magnitude of
the peculiar velocity, regardless of its orientation.

\section{Unsteady hydrodynamic behavior}

As is well known, hydrodynamics is one of the key properties of \emph{normal} fluids. Let us  imagine a  gas of \emph{elastic} particles in an arbitrary initial state defined by a certain distribution function $f^0(\mathbf{r},\mathbf{v})$. The standard evolution scenario starting from that initial state occurs along two consecutive stages \cite{DvB77}. First, during the so-called \textit{kinetic} stage, the velocity distribution function $f(\mathbf{r},\mathbf{v},t|f^0)$, which functionally depends on the initial state, experiments a quick relaxation (lasting of the order of a few collisions per particle) toward a  ``normal''  form where all the spatial and temporal dependence
takes place through a functional dependence on the hydrodynamic
fields $n$, $\mathbf{u}$, and $T$, i.e., $f(\mathbf{r},\mathbf{v},t|f^0)\to f[\mathbf{v}|n,\mathbf{u},T]$. Next, during the \textit{hydrodynamic} stage, a slower evolution occurs. While the first stage is very sensitive to the initial preparation of the system, the details of the initial
state are practically ``forgotten'' in the hydrodynamic regime.

An extremely important issue is whether or not the above two-stage scenario maintains its applicability in the inelastic case. For the sake of concreteness, let us consider the Boltzmann equation for a driven homogeneous granular gas (in the Lagrangian frame):
\beq
\partial_t f(\mathbf{V},t)-\omega \mathcal{F} f(\mathbf{V},t)=J[\mathbf{V}|f,f],
\label{3.1}
\eeq
where $\mathcal{F}$ is an (isotropic or anisotropic) operator representing the external driving and $\omega$ is a constant  measuring the strength of the driving. The operator is assumed to preserve total mass and momentum. The original problem can indeed be homogeneous \cite{MS00,GMT12,MGT12} or become equivalent to a homogeneous problem after a certain change of variables. The latter situation happens for the uniform shear flow (USF) \cite{AS12,DSBR86,AS05,AS07} and the ULF [see Eq.\ \eqref{2.3}].

Given an operator $\mathcal{F}$, the solution to Eq.\ \eqref{3.1} depends functionally on the initial distribution $f^0$ and parametrically on the value of the strength $\omega$. Since the only time-dependent hydrodynamic variable is the temperature $T(t)$, the existence of a hydrodynamic
regime implies that, after a certain number of collisions per
particle,
\beq
f(\mathbf{V},t|f^0,\omega)\to n\left[{m}/{2 T(t)}\right]^{3/2}
f^*(\mathbf{C}(t);\omega^*(t)),\quad \mathbf{C}(t)\equiv \frac{\mathbf{V}}{\sqrt{2T(t)/m}},\quad \omega^*(t)\equiv \frac{\omega}{K\left[T(t)\right]^\gamma}.
\label{1}
\eeq
Here, $\mathbf{C}$ is the
(peculiar) velocity in units of the (time-dependent) thermal speed and $\omega^*$ is the \emph{reduced} driving strength, where the choices of the constant $K$ and the exponent $\gamma$ are dictated in each case by dimensional analysis. The scaled velocity distribution
function $f^*(\mathbf{C};\omega^*)$ should be, for a given value of the
coefficient of restitution $\alpha$, a \emph{universal} function in
the sense that it is independent of the initial state $f^0$ and
depends on the driving strength $\omega$ through the reduced
quantity $\omega^*$ only. In other words, if a hydrodynamic description
is possible, the different solutions $f(\mathbf{V},t|f^0,\omega)$ of the
Boltzmann equation \eqref{3.1} would be ``attracted'' to the
universal form \eqref{1}. This has been confirmed by DSMC simulations for a stochastic white-noise driving ($\mathcal{F}=\partial_{\mathbf{V}}^2$) at the level of the cumulants $a_2$ and $a_3$ \cite{GMT12}, for the USF ($\mathcal{F}=V_y\partial_{V_x}$) at the level of the viscosity, the viscometric functions, and the marginal distributions \cite{AS12,AS07}, and for the ULF ($\mathcal{F}=\partial_{V_x}V_x$) at the level of the viscosity \cite{AS12}.The case of a Gauss' driving ($\mathcal{F}=- \partial_{\mathbf{V}}\cdot\VV$) \cite{MS00} is special in the sense that, once the hydrodynamic regime is reached,
$f^*(\mathbf{C}(t);\omega^*(t))$ is a constant function of $\omega^*(t)$ \cite{MGT12}.
We conjecture that Eq.\ \eqref{1} applies as well to the case of a combination of the  Gauss' and the stochastic white-noise drivings \cite{GSVP11a,GSVP11b}.

Translated to the ULF case, Eq.\ \eqref{1} implies that
\beq
\eta^*(\tau|\rho^0,a_0)\to \eta^*(a^*(\tau)),\quad a_{2,3}(\tau|\rho^0,a_0)\to a_{2,3}(a^*(\tau)),
\label{3.2}
\eeq
\beq
\rho_{x,y}(V_{x,y},\tau|\rho^0,a_0)\to \sqrt{{m}/{2T(\tau)}}g_{x,y}(C_{x,y}(\tau);a^*(\tau)),\quad
P(V,\tau|\rho^0,a_0)\to \sqrt{{m}/{2T(\tau)}}F(C(\tau);a^*(\tau)).
\label{3.3}
\eeq
As said above, the validity of the first term of Eq.\ \eqref{3.2} was addressed in Ref.\ \cite{AS12}.
In the next section we extend the analysis to $a_2(a^*)$, $a_3(a^*)$, $g_x(C_x;a^*)$, $g_y(C_y;a^*)$, and $F(C;a^*)$.

\section{Results}

We have numerically solved the Boltzmann equation \eqref{2.3} by the DSMC method for three values of the coefficient of restitution ($\alpha=0.5$, $0.7$, and $0.9$) and three values of the strain rate ($a_0/\nu_0(0)=-11.26$, $-0.011$, and $0.011$). The two negative values of $a_0$ correspond to a compressed ULF, so that the viscous heating  term $\frac{2}{3}|a_0| T_x$ competes with the cooling term $\zeta_0 T$ in the first equation of Eq.\ \eqref{energy_balance2}. The magnitude of  $a_0/\nu_0(0)=-11.26$ is large enough as to make the viscous heating initially prevail  over the inelastic cooling (even for $\alpha=0.5$). As the granular gas heats up, the cooling term grows more rapidly than the heating term until eventually both terms cancel each other and a steady state is reached. Conversely, the magnitude of  $a_0/\nu_0(0)=-0.011$ is so small that  the viscous heating is initially dominated by the inelastic cooling (even for $\alpha=0.9$) and the granular gas cools down. Now, the cooling term decays more rapidly than the heating term until the same steady state as before is eventually reached. On the other hand,  the positive value $a_0/\nu_0(0)=0.011$ corresponds to an expanded ULF and both terms  $\frac{2}{3}a_0 T_x$ and $\zeta_0 T$ produce a cooling effect. Therefore, $T(\tau)$ monotonically decreases (and thus $a^*(\tau)$ monotonically increases) without any bound and no steady state exists.

For each one of the nine pairs $(\alpha,a_0)$ we have considered five  initial conditions.
First, we have taken the local equilibrium state
\beq
\rho^0(\VV)=\left(\frac{m}{2\pi T^0}\right)^{3/2}e^{-mV^2/2T^0},
\label{4.5}
\eeq
where $T^0$ is the (arbitrary) initial temperature. The initial longitudinal temperature, cumulants and marginal distributions are simply
\beq
T_x(0)=T^0,\quad a_2(0)=a_3(0)=0,
\eeq
\beq
\rho_x(V_x,0)=\sqrt{\frac{m}{2\pi T^0}}e^{-mV_x^2/2T^0}
,\quad \rho_y(V_y,0)=\sqrt{\frac{m}{2\pi T^0}}e^{-mV_y^2/2T^0},\quad
P(V,0)=4\pi V^2 \left(\frac{m}{2\pi T^0}\right)^{3/2}e^{-mV^2/2T^0}.
\eeq
Besides, we have considered four anisotropic initial conditions of the  form
\beq
\rho^0(\VV)=\frac{1}{2}\sqrt{\frac{m}{2\pi T^0}}e^{-mV_z^2/2T^0}\left[\delta\left(V_x-V^0\cos
\phi\right)\delta\left(V_{y}+V^0
\sin \phi\right)+\delta\left(V_x+V^0\cos
\phi\right)\delta\left(V_{y}-V^0
\sin \phi\right)\right],
\label{4.6}
\eeq
where $V^0\equiv\sqrt{2T^0/m}$
is the initial thermal speed and $\phi=0$, $\pi/4$, $\pi/2$, and $3\pi/4$. In this case,
\beq
T_x(0)=2T^0\cos^2\phi,\quad a_2(0)=-\frac{4}{15},\quad a_3(0)=-\frac{32}{105},
\eeq
\beq
\rho_x(V_x,0)=\frac{1}{2}\left[\delta\left(V_x-V^0\cos
\phi\right)+\delta\left(V_x+V^0\cos
\phi\right)\right],\quad
\rho_y(V_y,0)=\left[\delta\left(V_{y}-V^0
\sin \phi\right)+\delta\left(V_{y}+V^0
\sin \phi\right)\right],
\eeq
\beq
P(V,0)=\sqrt{\frac{m}{2\pi T^0}}e^{-(mV^2/2T^0-1)} \frac{2V}{\sqrt{V^2-2T^0/m}}\Theta(V^2-2T^0/m),
\label{P0}
\eeq
where $\Theta$ is the Heaviside step function.

In the course of the simulations we have measured $T(\tau)$, $T_x(\tau)$, $a_2(\tau)$, $a_3(\tau)$, $\rho_x(V_x,\tau)$, $\rho_y(V_y,\tau)$, and $P(V,\tau)$ for each one of the 45 cases described above. {}From $T(\tau)$ and $T_x(\tau)$ the temporal evolution of the reduced strain rate $a^*(\tau)$ [cf.\ Eq.\ \eqref{longitudinal_rate}] and the reduced viscosity $\eta^*(\tau)$ [cf.\ Eq.\ \eqref{eta}] has been followed. Elimination of time between both quantities allows one to get a parametric plot of $\eta^*$ versus $a^*$. In Ref.\ \cite{AS12} we observed that, after a few collisions per particle, the  curves corresponding to the five initial conditions for each one of the nine values of the pair $(\alpha,a_0)$ collapse to a common \emph{hydrodynamic} curve. For instance, in the case $\alpha=0.5$ the duration of the kinetic stage was about $3$--$4$ collisions per particle for $a_0/\nu_0(0)=-11.26$ and about $7$--$8$ collisions per particle for $a_0/\nu_0(0)=\pm 0.011$, while the total relaxation period toward the steady-state values $a_s^*$ and $\eta_s^*$ took about 20 collisions per particle for $a_0/\nu_0(0)=-11.26$ and $a_0/\nu_0(0)=- 0.011$.

\begin{figure}
\scalebox{1}[1]{\includegraphics{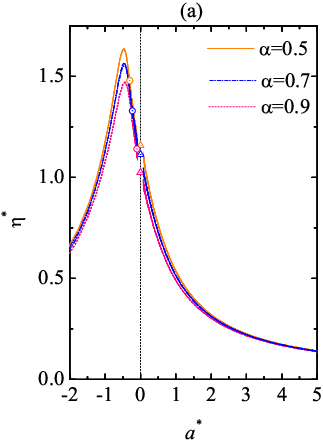}\includegraphics{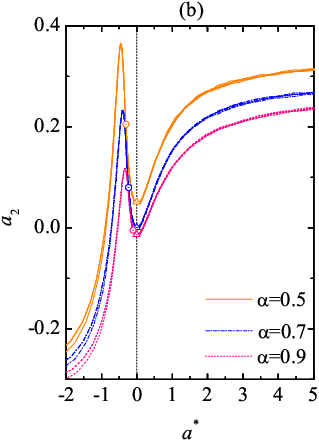}\includegraphics{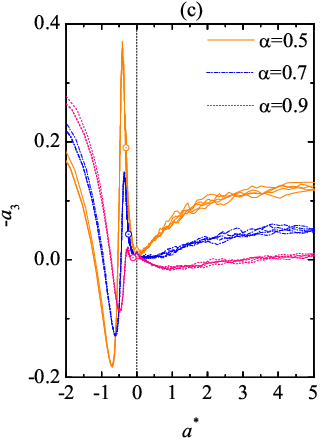}}
\caption{Plot of (a) the reduced viscosity $\eta^*$, (b) the second cumulant $a_2$, and (c) the third cumulant $a_3$
versus the reduced strain rate $a^*$ for  $\alpha=0.5$ (orange solid lines), $\alpha=0.7$ (blue dash-dotted lines), and $\alpha=0.9$ (pink dotted lines).  The circles represent the steady-state points, while the triangles represent the values in the HCS.
}
\label{fig_eta}
\end{figure}

\begin{figure}
\scalebox{.95}[.95]{\includegraphics{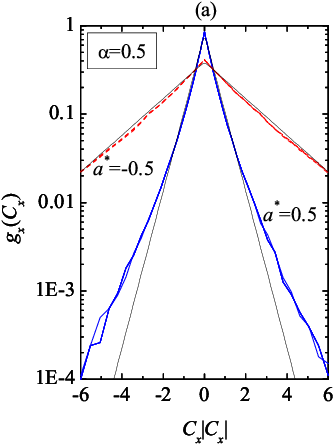}\includegraphics{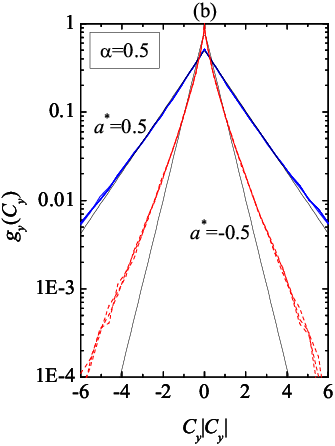}\includegraphics{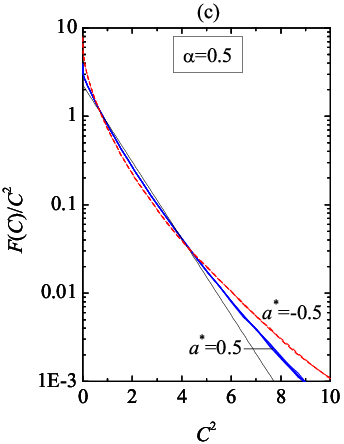}}
\caption{Marginal probability distribution functions (a) $g_x(C_x)$, (b) $g_y(C_y)$, and (c) $F(C)$
for $\alpha=0.5$ at $a^*=-0.5$ (red dashed lines) and $a^*=0.5$ (blue solid lines). The black  thin solid lines represent the corresponding Maxwellian distributions.
}
\label{fig_g}
\end{figure}

\begin{figure}[t]
\scalebox{1.0}[1.0]{\includegraphics{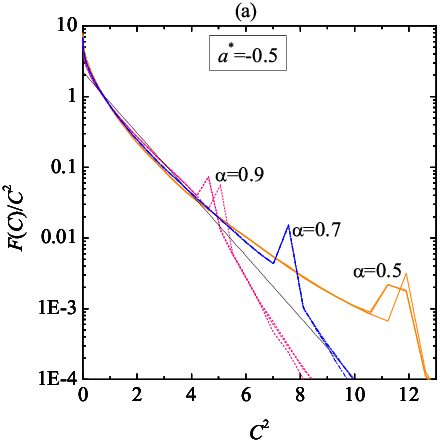}\hspace{1cm}\includegraphics{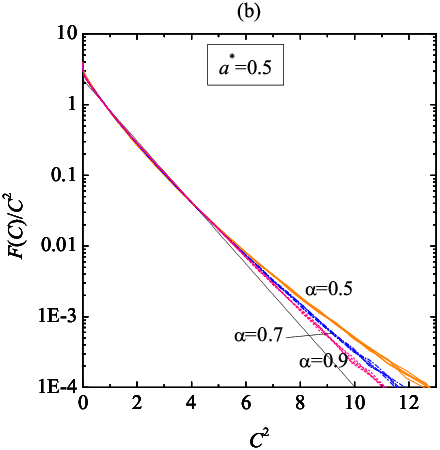}}
\caption{Marginal probability distribution function  $F(C)$ for $\alpha=0.5$ (orange solid lines), $\alpha=0.7$ (blue dash-dotted lines), and $\alpha=0.9$ (pink dotted lines) at (a) $a^*=-0.5$ and (b) $a^*=0.5$. The black  thin solid lines represent the Maxwellian distributions.
}
\label{fig_F}
\end{figure}

Although the non-Newtonian viscosity in the  ULF was analyzed in Ref.\ \cite{AS12}, for the sake of completeness we show in Fig.\ \ref{fig_eta}(a) $\eta^*(a^*)$ for three windows of $a^*$ where the hydrodynamic regime is practically established: $-2\leq a^*\leq a_s^*$ (corresponding to $a_0/\nu_0(0)=-11.26$), $-0.08\geq a^*\geq a_s^*$ (corresponding to $a_0/\nu_0(0)=-0.011$), and $ a^*\geq 0.08$ (corresponding to $a_0/\nu_0(0)=0.011$). A non-monotonic behavior of $\eta^*(a^*)$ is observed, with a maximum at $a^*=-0.47$ ($\alpha=0.5$ and $0.7$) and $a^*=-0.44$ ($\alpha=0.9$).
In the figure the open circles represent the steady state points $(a_s^*,\eta_s^*)$ and the open triangles at $a^*=0$ represent values of the NS viscosity  obtained independently \cite{BRC99,BR04,BRMG05,MSG05,GSM07}. While, for each $\alpha$, the steady-state point is an attractor of the heating ($a^*<a^*_s$) and cooling ($a^*>a^*_s$) branches with negative $a^*$, the NS point is a repeller of the two cooling branches ($a_s^*<a^*<0$ and $a^*>0$) \cite{S08a,S09}. It is noteworthy that, although the need of discarding the kinetic stage creates the gap $-0.08\geq a^*\geq 0.08$, the extrapolations of the cooling branches with positive and negative $a^*$ smoothly join at the NS point.

In the case of the heating states ($a_0/\nu_0(0)=-11.26$) the evolution is so fast when starting from  the local equilibrium initial condition \eqref{4.5} and from the anisotropic initial condition \eqref{4.6} with $\phi=\pi/2$ that the corresponding curves join the hydrodynamic line past the maximum located at $a^*<a_s^*$. Thus, in Fig.\ \ref{fig_eta} and henceforth those two initial conditions are omitted in the case $a_0/\nu_0(0)=-11.26$.

Figures \ref{fig_eta}(b) and \ref{fig_eta}(c) show the cumulants $a_2(a^*)$ and $a_3(a^*)$, respectively, for the same windows of $a^*$ as in Fig.\ \ref{fig_eta}(a). Again, the open circles represent the steady-state points.
The open triangles at $a^*=0$ correspond to homogeneous cooling state (HCS) values obtained independently \cite{MS00,BP06,BP06b,SM09}. We observe that, for each $\alpha$, the HCS values are fully consistent with the extrapolation to $a^*\to 0$ of the two cooling branches.
For these higher-order moments, the overlapping of the individual heating curves in the region $-2\leq a^*\leq -1$ is less robust than in the case of $\eta^*$. As indicated before, this is a consequence of the very fast evolution taking place for the states with  $a_0/\nu_0(0)=-11.26$. In fact, the value $a^*=-2$ is reached after less than $0.3$ collisions per particle only. We have checked that starting from more negative values of $a_0/\nu_0(0)$ hardly changes the situation. We also observe that, for each $\alpha$, the noise in the cooling curves with positive $a^*$ increases as the order of the moment  grows.

The dependence of $a_2$  on $a^*$ is similar for the three values of $\alpha$. The second cumulant is negative at $a^*=-2$, grows with increasing $a^*$, changes sign, and reaches a maximum value at $a^*=-0.45$ ($\alpha=0.5$), $a^*=-0.40$ ($\alpha=0.7$), and $a^*=-0.33$ ($\alpha=0.9$). Thereafter, $a_2$ decays, reaches a local minimum at $a^*=0$ and then grows with increasing positive $a^*$. At a fixed value of $a^*$, we observe that $a_2$ increases with increasing inelasticity.
The dependence of $a_3$ on $a^*$ for compression states ($a^*<0$) is more complex than that of $a_2$. Instead of a maximum, $-a_3(a^*)$ presents a minimum at $a^*=-0.70$ ($\alpha=0.5$), $a^*=-0.61$ ($\alpha=0.7$), and $a^*=-0.48$ ($\alpha=0.9$), followed by a maximum at $a^*=-0.41$ ($\alpha=0.5$), $a^*=-0.36$ ($\alpha=0.7$), and $a^*=-0.25$ ($\alpha=0.9$). Moreover, the curves corresponding to the different values of $\alpha$ cross each other in the region of negative $a^*$.

Now we turn our attention to the marginal distributions $g_x(C_x;a^*)$, $g_y(C_y;a^*)$, and $F(C;a^*)$ [cf.\ Eq.\ \eqref{3.3}]. Although we have evaluated those functions for the whole temporal evolution of the systems, here we focus on two representative instantaneous values of the reduced strain rate: $a^*=-0.5$ and $a^*=0.5$. The first value belongs to the heating compression branch and is reached after slightly less than $2$ collisions per particle. The second value belongs to the cooling expansion branch and is reached after $13$ ($\alpha=0.5$), $18$ ($\alpha=0.7$), or $38$ ($\alpha=0.9$) collisions per particle.
The functions $g_x$ and $g_y$ for $\alpha=0.5$ are plotted in Figs.\ \ref{fig_g}(a) and \ref{fig_g}(b), respectively. The representation in the horizontal and vertical axes is chosen to visualize deviations from the  Maxwellians $g_{x,y}^M(C_{x,y})=(\pi T_{x,y}/T)^{-1/2}\exp(-C_{x,y}^2T/T_{x,y})$. We first note a high degree of collapse   of data corresponding to different initial conditions to a common curve in the velocity domain $C_{x,y}^2<6$.
Comparison of $g_x(C_x)$ for $a^*=-0.5$ and $a^*=0.5$ shows a much broader distribution in the first case than in the second one. The opposite happens for $g_y(C_y)$. This is consistent with the measured values $T_x/T= 2.09$ and $T_y/T=0.45$ at $a^*=-0.5$ and $T_x/T= 0.49$ and $T_y/T=1.26$ at $a^*=0.5$. Apart from that, we observe an overpopulated high-velocity tail with respect to the Maxwellian $g_{x,y}^M(C_{x,y})$. In the case of the broader distributions this overpopulation phenomenon occurs beyond the region $C_{x,y}^2<6$ displayed in Figs.\ \ref{fig_g}(a) and \ref{fig_g}(b).

The probability distribution function for the reduced speed, $F(C)$, is plotted in Fig.\ \ref{fig_g}(c) for $\alpha=0.5$. Again, the representation is chosen to highlight deviations from the Maxwellian $F^M(C)=4\pi^{-1/2}C^2e^{-C^2}$. We see that $F(C)$ for both the compression ($a^*=-0.5$) and the expansion ($a^*=0.5$) states exhibits strong departures from the Maxwellian $F^M(C)$, with overpopulated tails. This is especially so in the case of $a^*=-0.5$, in agreement with the fact that $a_2$ and $|a_3|$ are clearly larger at $a^*=-0.5$ than at at $a^*=0.5$ [cf.\ Figs.\ \ref{fig_eta}(b) and \ref{fig_eta}(c)].

The function $F(C)$ is plotted in Fig.\ \ref{fig_F} for the three values of the coefficient of restitution and the  two chosen values of the reduced strain rate.
The main observation is that $F(C)$ for $a^*=-0.5$ presents a singular behavior at $C^2\approx 12$, $7.5$, and $5$ for $\alpha=0.5$, $0.7$ and $0.9$, respectively. This is a remaining artifact associated with the special initial conditions \eqref{4.6}. As shown by Eq.\  \eqref{P0}, at $t=0$ all the particles have a speed $V>\sqrt{2T^0/m}$ and the distribution function diverges at $V=\sqrt{2T^0/m}$. Since $a^*=-0.5$ is reached after not more than 2 collisions per particle, there exists  a certain population of surviving particles which have not collided yet. Those particles are subject to the heating effect due to the non-conservative external force $-m a_0V_x
\widehat{\bf x}$ but not to the collisional cooling effect. Consequently, they increase their energy more than the rest of the particles, have an ever increasing reduced speed $C$, and thus contribute to the tail of $F(C)$ only. As the inelasticity decreases the cooling effect becomes less important and the reduced speed of the surviving particles grows more slowly.
Therefore, in the case $a^*=-0.5$, the \emph{tail} of the distribution function $F(C)$ for values of $C$ equal to or larger than the singularity is still dependent on the initial state and cannot be considered as hydrodynamic yet. On the other hand, the tail does not have any practical influence on the first few velocity moments.
In the case $a^*=0.5$ the number of collisions is much higher and thus the influence of the surviving particles is negligible.

\section{Conclusions}
In this paper  we have presented DSMC numerical solutions of the inelastic Boltzmann equation for the unsteady ULF in the co-moving Lagrangian frame. In order to uncover the three types of possible regimes (heating compression states, cooling compression states, and cooling expansion states), three values of the initial strain rate have been considered. Starting from different initial conditions, the temporal evolution of the granular temperature $T$, the longitudinal temperature $T_x$, the velocity cumulants $a_2$ and $a_3$, and the marginal probability distribution functions $\rho_x(V_x)$, $\rho_y(V_y)$, and $P(V)$ has been recorded. By eliminating time in favor of the reduced strain rate $a^*$ we have checked that, after a first kinetic stage lasting a few collisions per particle, the curves corresponding to different initial states tend to collapse to  common ``hydrodynamic'' curves. These findings extend to the cumulants and to the distribution function recent results \cite{AS12} obtained for the non-Newtonian viscosity.
The dependence of the cumulants $a_2$ and $a_3$ on $a^*$ exhibits a non-monotonic behavior with interesting features. In consistency with this, the velocity distribution functions are seen to  depart from a Maxwellian form.


\begin{theacknowledgments}

This work has been supported by  the Spanish Government through Grant No.\ FIS2010-16587 and by the Junta de Extremadura (Spain) through Grant No.\ GR10158, partially financed by FEDER  funds.
\end{theacknowledgments}



\bibliographystyle{aipproc}   

\bibliography{D:/Dropbox/Public/bib_files/Granular}

\end{document}